\documentstyle[12pt]{article}

\newcommand{\nc}{\newcommand}
\nc{\beq}{\begin{equation}}
\nc{\eeq}{\end{equation}}
\nc{\beqa}{\begin{eqnarray}}
\nc{\eeqa}{\end{eqnarray}}

\input epsf
\newwrite\ffile\global\newcount\figno \global\figno=1

\def\writedef#1{}
\def\figin{\epsfcheck\figin}\def\figins{\epsfcheck\figins}
\def\epsfcheck{\ifx\epsfbox\UnDeFiNeD
\message{(NO epsf.tex, FIGURES WILL BE IGNORED)}
\gdef\figin##1{\vskip2in}\gdef\figins##1{\hskip.5in}
\else\message{(FIGURES WILL BE INCLUDED)}%
\gdef\figin##1{##1}\gdef\figins##1{##1}\fi}
\def\figinsert{}
\def\ifig#1#2#3{\xdef#1{fig.~\the\figno}
\writedef{#1\leftbracket fig.\noexpand~\the\figno}%
\figinsert\figin{\centerline{#3}}\medskip\centerline{\vbox{\baselineskip12pt
\advance\hsize by -1truein\center\footnotesize{  Fig.~\the\figno.} #2}}
\bigskip\endinsert\global\advance\figno by1}
\def\endinsert{}

\begin{document}

\title{\large{\bf Effective Field Theory of Neutron Star Superfluidity}}

\author{
James Hormuzdiar\thanks{james.hormuzdiar@yale.edu} \\
Department of Physics and Astronomy, \\
Yale University, New Haven, CT 06520 \\ \\
Stephen D.H.~Hsu\thanks{hsu@duende.uoregon.edu} \\
Department of Physics, \\
University of Oregon, Eugene OR 97403-5203 \\ \\   }

\date{October, 1998}

\maketitle

\begin{picture}(0,0)(0,0)
\put(350,360){OITS-665}
\put(350,340){YCTP-P25-98}
\end{picture}
\vspace{-24pt}

\begin{abstract}
We apply effective field theory and renormalization group techniques to the 
problem of Cooper pair formation in neutron stars. Simple analytical 
expressions for the $^1 S_0$ condensate are derived which are free of 
nuclear potential model dependencies. The condensate is evaluated using 
phase shift data from neutron-neutron scattering.

\end{abstract}

\newpage


Neutron superfluidity plays an important role in the physics of neutron stars,
affecting the neutrino cooling rate and heat capacity as well as starquake 
phenomena  \cite{Nstar}. However, the computation of the neutron-neutron 
(NN) condensate is a difficult problem and has resulted in a wide range of 
results \cite{Wam,Clark,Schulze,barep,Khodel}. 
The main reason is the exponential sensitivity of the 
condensate to the effective interaction, which is typically extracted in a 
model-dependent way from NN scattering data, and also depends on medium 
effects. 

In this paper, we use the renormalization group (RG) to construct the 
effective field theory of neutrons near the Fermi surface. In this description 
the size of the condensate is related to the position of a Landau pole in a 
running coupling near the Fermi surface (FS) \cite{FS,EHS}. All interactions 
other than the Cooper pairing interaction (a four-neutron operator 
restricted to 
kinematic points describing scattering of neutrons with equal and opposite 
momenta) can be shown to be irrelevant at the FS. Due to this simplification 
the evolution of the Cooper pairing interaction, and hence the location
of the Landau pole, can be determined analytically. Medium effects are 
reflected in the RG evolution of Fermi liquid parameters such as the
effective mass and four-neutron operator, 
whose initial form $G(p^2)$ 
is obtained by matching to NN phase shifts via an exact expression for the 
in-medium NN amplitude. 
We focus on the $^1S_0$ condensate, although our techniques can 
also be applied to study condensates of higher angular momentum.

We begin by reviewing the effective field theory description of Fermi 
liquids \cite{FS,EHS}.
In this description we make a guess as to the form of the
effective theory close to the Fermi surface. The obvious guess based on
the dynamics of non-relativistic systems is that the theory is one of
weakly interacting fermions: these are the dressed ``quasi-particles''
of solid state physics language. We will henceforth refer to these effective 
degrees of freedom as ``neutrons'', with the understanding that they could
in principle be related to the bare neutrons in a complicated way.
Rather than treating other degrees of freedom such as pions, deltas, etc.,  
as propagating degrees of freedom we will integrate them out leaving a
potentially infinite sum over non-local, higher dimension fermion operators. 
The effective Lagrangian is simply
\beq
\label{lagrangian}
{\cal L } = 
\bar{\psi}_i (  i \partial\!\!\!/ ~+~ \mu  \gamma_0 ~-~ m ) 
\psi_i ~ +  ~\cdots~,
\eeq
where the ellipsis denote higher dimension interaction terms.
Although we will eventually specialize to the non-relativistic limit, 
we begin with a relativistic formulation because it allows us to 
systematically track the corrections to that limit.

The chemical potential in (\ref{lagrangian}) naturally breaks the O(3,1)
invariance of space-time to O(3) and furthermore picks out momenta of
order $p_F = \sqrt{\mu^2 - m^2}$. 
It is therefore natural to study the theory as we approach
the Fermi surface in a Wilsonian sense. We parameterize four momenta in the
following fashion
\beq 
p^{\mu} = (p^0, \vec{p}) = (k^0, \vec{k} + \vec{l})
\eeq
where $\vec{k}$ lies on the Fermi surface ($|\vec{k}| = p_F$) 
and $\vec{l}$ is orthogonal to it. 
We study the Wilsonian effective theory of modes near the Fermi surface,
with energy and momenta  
\beq
\label{FSE}
|k_0|,|\vec{l}| ~<~ \Lambda~~~~,~~~~ \Lambda \rightarrow 0~~~.
\eeq  
While this type of RG scaling is somewhat
unfamiliar, it actually corresponds to thinning out fermionic degrees
of freedom according to their eigenvalues under the operator
$i \partial\!\!\!/  + \mu \gamma_0 - m$. 
It is easy to see that eigenspinors of this operator with
eigenvalues $~\lambda_n:~ | \lambda_n | < \Lambda~$ correspond to states 
satisfying (\ref{FSE}). Consider an eigenspinor of the form
\beq
\psi_p ~=~ u(E,\vec{p})~ e^{ i( p \cdot x - p_0 t)}~~~,
\eeq
where $u(E,\vec{p})$ satisfies the usual momentum-space Dirac equation with
$E = \sqrt{ p^2 + m^2} = m + \mu ~\pm~ {\cal O} (\Lambda)$, and 
$p_0 = m \pm {\cal O} ( \Lambda )$. Then by direct substitution
we see that $\psi_p$ satisfies 
\beq
( i \partial\!\!\!/  ~+~ \mu \gamma_0  ~-~ m )~ \psi_k ~=~ 
{\cal O} ( \Lambda )~~~.
\eeq
Thus, the RG flow towards the Fermi surface just corresponds to taking 
the cutoff on eigenvalues of $i \partial\!\!\!/  + \mu \gamma_0 - m$ to zero.

Which operators are relevant in this limit? 
For our guess to make sense the kinetic term  
for the fermions must be invariant  when we
scale energies and momenta,
$k_0 \rightarrow s k_0$ (or, $t \rightarrow t/s$), 
and $\vec{l} \rightarrow s \vec{l}$,
 with $s < 1$. We must be careful to
satisfy all the global symmetries of the theory. In particular, there is
a spurious symmetry of (\ref{lagrangian}) in which we treat $\mu$ as
the temporal component of a fictitious $U(1)_B$ gauge boson. 
In other words, the combined transformations 
$\psi \rightarrow e^{i \theta t}~ \psi$ and $\mu \rightarrow \mu + \theta$ 
leave the lagrangian (\ref{lagrangian}) invariant. From this, we deduce
that time derivatives acting on $\psi$ and factors of $\mu$ may only 
enter the effective theory in the combination
$( \partial_0  + \mu ) \gamma_0$. This requires the kinetic term of our
effective theory to be of the form
\beq
\label{KE}
{\cal S}_{eff} = \int dt~ d^3p ~ \bar{\psi} \,
\Big( (i \partial_t   + \mu) \gamma_0 ~-~ \vec{p}\cdot \vec{\gamma}
~-~  m \Big) \, \psi ~~,
\eeq
where $m$ is the effective neutron mass.
In the Wilsonian RG scaling, we eliminate all modes with
energy and momenta  $|k_0|,|l| > \Lambda$, where $\Lambda$ is our cutoff. 
As discussed above, on the remaining degrees of freedom the operator 
$(i \partial_t  + \mu) \gamma_0 - \vec{p}\cdot \vec{\gamma} - m ~
\sim {\cal O}(\Lambda)$ and therefore scales like $s$.
We deduce that for (\ref{KE}) to remain invariant,  
$\psi$ must scale as $s^{-1/2}$. 
Now consider the four fermion operator
\beq
\label{ffop}
G \int dt\, d^3\vec{p_1}\,d^3\vec{p_2}\,d^3\vec{p_3}\,d^3\vec{p_4}  ~
\bar{\psi} \Gamma \psi ~ \bar{\psi} \Gamma \psi
~\delta^3 ( \vec{p}_1 + \vec{p}_2 + \vec{p}_3 + \vec{p}_4 )
\eeq
where $\Gamma$ contains any Lorentz or flavor structure. 
Naively, for $\vec{l}$ close to zero, the delta function does not scale, 
and (\ref{ffop}) is irrelevant since it scales as $s$.
Higher dimension operators with extra powers of the
fields are clearly irrelevant as well. 
The only operators
that survive are four-fermion operators satisfying the kinematic constraint 
$\vec{p}_1 \simeq - \vec{p}_2$, in which case the
delta function becomes
\beq
\delta(\vec{l_1} + \vec{l_2} - \vec{l_3} - \vec{l_4})
\eeq
and scales as $s^{-1}$. The resulting four fermion operator is marginal,
and quantum effects must be considered to determined its relevance to
dynamics at the FS.


We now study the RG flow of the marginal Cooper pairing
operator. We are primarily interested in the $^1 S_0$ component of
this operator, which will be seen to dominate the others near the FS.
We first consider the $^1 S_0$ component by itself, and later consider
corrections to this approximation that result from loops involving 
higher angular momentum components. It is easy to show that angular momentum
conservation forbids such corrections so the $^1 S_0$ evolution is exact.

Let us briefly review the kinematics of neutron-neutron scattering at the
Fermi surface in the center of mass frame. Let the incoming momentum be
$\vec{p}_1$, and the outgoing momentum $\vec{p}_2$, both of magnitude $p_F$. 
The scattering amplitude is in general a function of the angle between 
these two vectors. The form factor of an  operator describing this 
scattering process can be decomposed into components corresponding to
projections onto different angular momentum eigenstates, or spherical
harmonics $Y^l_m (\theta, \phi)$. Actually, since the amplitudes are
independent of $\phi$ only the $m = 0$ components are required, and the
projection just involves integration over $\theta$. The $^1S_0$ component
is obviously the component which is independent of $\theta$, and hence has
a constant form factor.

\epsfysize=6 cm
\begin{figure}[htb]
\center{
\leavevmode
\epsfbox{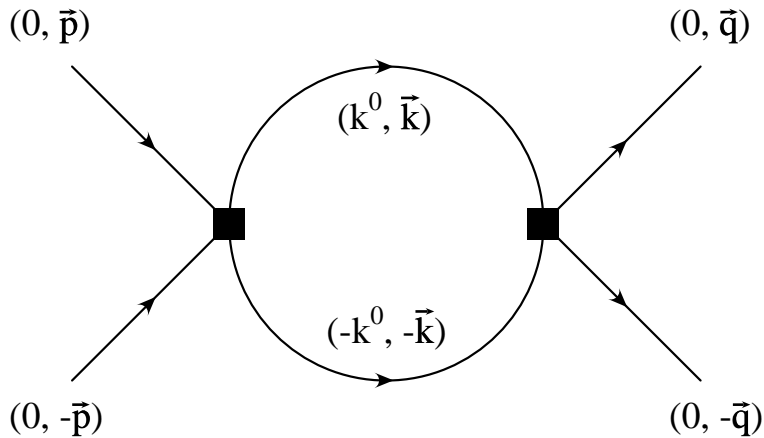}
\caption{One loop renormalization of G} \label{bubble}
}
\end{figure}

The beta function for our four-neutron operator can be deduced from
the one loop graph in figure \ref{bubble}. If we consider only the $^1S_0$ 
component of the coupling the powers of $G$ can be factored out of
the integral, yielding\footnote{Note that the $i \epsilon$ prescription for propagators
at finite chemical potential is slightly subtle \cite{Polonyi}. Essentially, the
sign of the $i \epsilon$ terms is such that the usual Wick rotation can be made
regardless of the sign of $( E( \vec{p}) - \mu )$.}  
\beq
-G^2\int {d^4 k \over(2 \pi)^4} 
\left[ {i \over k^\mu \gamma_\mu + \mu \gamma_0 - m} \right]_{i k}
\left[ {i \over -k^\mu \gamma_\mu + \mu \gamma_0 - m} \right]_{j l}~~~.
\eeq
We can rewrite this as
\beq
\label{dq}
G^2\left[-{(\mu \gamma_0 + m)_{ik} \over 2} {(\mu \gamma_0 + m)_{jl} \over 2}
+ {1 \over 12 } p_F^2 
\sum_{a=1}^3 (\gamma_a)_{ik}(\gamma_a)_{jl}\right]I~~~,
\eeq
where the integral $I$ is given by
\beqa
\label{I1}
I ~&=&~ {p_F \over \mu}\int {d \Omega_k \over (2 \pi)^3} \int {dk_0 dl \over 2 \pi}
{1 \over k_0^2 - l^2 - i \epsilon} \nonumber \\
&=&~  {i ~ p_F \over 2 \pi^2 \mu}
\ln\left( {\Lambda_{IR}\over \Lambda_{UV}} \right)~~~.
\eeqa
Here $\Lambda_{IR}$ and $\Lambda_{UV}$ are the infrared and ultraviolet
limits of integration. In the usual Wilsonian sense, the effects of modes
in the shell between these cutoffs is summarized in the evolution of the
coupling $G$.
In the non-relativistic limit, where $\frac{p_F^2}{m^2} \rightarrow 0$, 
(\ref{dq}) becomes
\beq
\delta G = 
G^2\left[{\gamma_0+1 \over 2}\right]_{ik}
\left[{\gamma_0+1 \over 2}\right]_{jl} ~
{m~p_F \over 2 \pi^2} ~ t
\eeq
where $t = \ln({\Lambda_{IR} \over \Lambda_{UV}})$.
To incorporate non-relativistic corrections to some order in 
$p_F^2 / m^2$, one must include additional operators
appearing in (\ref{I1}) in the RG equations. 
However, in our case of interest, $p_F^2 / m^2$ is at most of order a 
few percent, so we will drop all non-relativistic corrections. 
Since ${1 \over 2} ( \gamma_0 \pm 1 )$ is simply the 
neutron/antineutron projection operator, the result only
renormalizes the coupling between neutrons, and does not involve
anti-neutrons. Henceforth we will focus on this
interaction. The resulting RG equation is
\beq
\label{RG0}
{d G \over dt} = {m~p_F \over 2 \pi^2}  G^2~~ ~.
\eeq
The solution to this equation is 
\beq
G (t) = {2 \pi^2 G (0) \over 2 \pi^2 - t ~m  p_F  G (0)}~,
\eeq
which has a Landau pole at 
\beq
t ~=~ {2 \pi^2 \over m  p_F  G (0)}~~~.
\eeq
By dimensional analysis, we expect a Cooper pairing gap of size
\beq 
\label{Del}
\Delta \simeq  \Lambda_{UV} e^{2 \pi^2 \over m p_F  G (0)}~~~.
\eeq
The exponent in (\ref{Del}) is very similar to the usual
BCS weak-coupling result \cite{Wam}, except that in our case the
coupling $G$ has a well-defined origin: it arises from the matching
of our purely nucleonic effective theory to the full theory and 
can be extracted from NN
scattering data. The prefactor will be determined by matching to known
results for the gap at low density.

Now we return to the issue of higher angular momentum components. As
discussed previously we work in the basis provided by the spherical harmonics 
$Y^l_m ( \theta, \phi )$. Actually we only require the $m=0$ harmonics,
which are $\phi$ independent.  
Breaking $G(\theta)$ into its spherical harmonic components
\beq
G(\theta) = G^{(0)} ~ Y^0_0(\theta, 0) ~+~ G^{(1)} ~ Y^1_0(\theta, 0) ~+~ 
G^{(2)} ~ Y^2_0(\theta, 0) ~+~ \cdots
\label {G-breakup}
\eeq
and repeating the previous analysis, the only change is in the
replacement
\beq
G^2 \int d^2 k \rightarrow \int d^2 k ~ G(\beta_{p k}) G(\beta_{k q})
\eeq
where $\beta_{a b}$ is the angle between $\vec a$ and $\vec b$.

Inserting (\ref{G-breakup}) and simplifying yields the new set of RG equations
\beq
\label{fullRGE}
{d G^{(l)} \over d t} 
= m \, p_F \sum_{{l'}=0}^{\infty} (G^{(l')})^2 \chi^{l {l'}}
\eeq
where $\chi^{ll'}$ is given in terms of integrals over Legendre polynomials.
It is easy to show that $\chi^{l l'}$ is diagonal, using the following result:
\beq
\int d^2k~ Y_0^l ( \beta_{pk},0 )~ Y_0^{l'} (\beta_{kq},0) 
~\propto~ Y_0^l ( \beta_{pq}, 0)~\delta^{l l'}~~~.
\eeq
This implies that there is no mixing between different angular momentum
components, and our treatment of the $^1S_0$ RGE is exact.


The final step in our analysis is to match the effective neutron Lagrangian
to the full nuclear theory, which must include pions as well as 
neutron-neutron contact terms \cite{KSW}. However, since our FS 
effective theory includes only very low energy quasiparticles, the 
pions have been integrated out. We need only retain the non-local 
four-neutron operator, whose 
value can be fixed by comparison with $^1S_0$ NN scattering data.

\epsfysize=2.4 cm
\begin{figure}[htb]
\center{
\leavevmode
\epsfbox{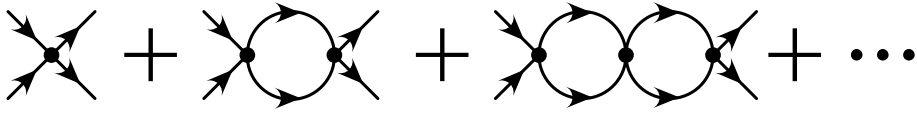}
\caption{Bubble Sum leading to $i \cal{M}$ } \label{bubblechain}
}
\end{figure}

First we note that the $^1S_0$ scattering amplitude in the theory
with only neutrons can be found by summing a bubble chain of Feynman
graphs (figure \ref{bubblechain}) 
in which the vertices are given by the Lagrangian form factor
$G (p^2, \nu)$. Here we work in the center of mass frame, so
$G$ is only a function of the neutron momentum $p$. We perform our
calculation in an effective theory living
in a thin shell around the Fermi surface, regulated by 
a hard ultraviolet cutoff $\Lambda_{UV}$ and a 
hard infrared cutoff $\Lambda_{IR}$. 
The resulting amplitude can be directly computed and is given by
\beq
\label{amp}
i {\cal M} = {-i G(p^2, \nu) \over 1+m ~ G(p^2, \nu) ~ (\nu + i p)/4 \pi}~~~,
\eeq
where 
\beq
\nu = {2 p \over \pi} \ln( {\Lambda_{IR} \over \Lambda_{UV}} ) ~~~.
\eeq
Invariance of the physical scattering amplitude ${\cal M}$ under changes
in $\nu$ implies the following RGE for the coupling $G(p^2,\nu)$:
\beq
\label{nuRG}
{1 \over G(p^2, \nu')} ~=~ {1 \over G(p^2, \nu)} ~+~ {m (\nu - \nu') \over
4 \pi}~~~,
\eeq
and a corresponding Landau pole at 
\beq
\nu_* ~=~ \nu ~+~ { 4 \pi \over m G(p^2, \nu)}~~~.
\eeq  
Comparison with (\ref{RG0}) reveals that the two RGEs are
identical.
The scale associated with the Landau pole is
\beq
\label{LP}
\Lambda_{IR}^* ~=~ \Lambda_{UV} e^{\pi \nu_* / 2 p}~~~.
\eeq
Note that, unlike the scattering amplitude ${\cal M}$,
the coupling $G(p^2, \nu)$ is not a physical quantity and cannot
be determined without fixing a subtraction scheme (i.e. specifying
$\nu$). This leads to an ambiguity in the overall normalization of
the scale associated with the Landau pole.

\begin{figure}[htb]
\center{
\leavevmode
\epsfbox{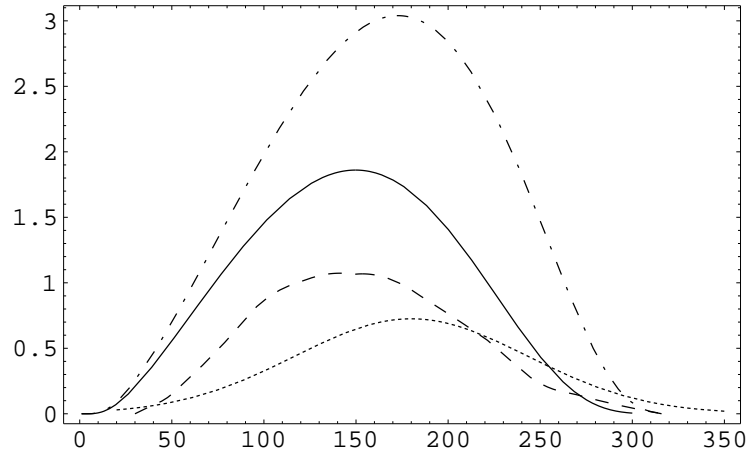}
\caption{Delta vs. Fermi momentum $p_F$. 
All units are MeV. Solid curve represents
Eq. (\ref{gapdelta}). Long-dashed and dot-dashed curves are from \cite{Wam}
and \cite{Khodel}. Short-dashed curve is (\ref{gapdelta}) with inputs from
\cite{Wam}.}
\label{Deltafig}
}
\end{figure}

We can rewrite (\ref{amp}) in terms of a phase shift, defined in terms of the
S-matrix by $S = e^{2 i \delta}$:
\beq
\delta = {1 \over 2 i} ~ \ln \left(1+i ~ {m ~ p \over 2 \pi} ~ 
{\cal M}\right)~~~.
\eeq
Finally, we can invert this relationship to obtain an expression for
the coupling,
\beq
G(p^2, \nu ) = -{4 \pi / m \over \nu + p ~ \cot(\delta)}~~~.
\eeq
Substituting this into (\ref{LP}) yields an equation relating
the superfluid gap (which must be equal to the scale of the Landau
pole, up to a some factor of order one) to the phase shift, 
\beq
\label{gapdelta0}
\Delta ~=~ \Lambda ~ e^{-{\pi \over 2} \cot\delta } ~~~ .
\eeq
The constant $\Lambda$ is undetermined due to the ambiguity mentioned above,
but scales like the Fermi energy $\epsilon_F = p_F^2 / 2m^*$, since it is
proportional to the UV cutoff or FS shell thickness. The
precise numerical value of the coefficient can be determined by studying 
the weak coupling limit of a low-density neutron gas ($p_F \rightarrow 0$).
Once determined, the coefficient remains fixed independent of the
Fermi momentum $p_F$. An explicit computation using the gap equation
has been performed in this limit by Khodel et al. \cite{Khodel}, with
the result $\Lambda = {8 \over e^2}\epsilon_F$.  The final result is
\beq
\label{gapdelta}
\Delta ~=~ {8 \over e^2} ~ {p_F^2 \over 2 m_*} ~ e^{-{\pi \over 2} \cot\delta } ~~~ .
\eeq
Our calculation can be viewed as a rigorous justification for the use
of a BCS-like expression depending on Fermi liquid 
parameters which incorporate in-medium
effects \cite{Wam}. In our case the coefficent of the exponential can
be justified from first principles. 
Equation (\ref {gapdelta}) is exact when evaluated using the
effective mass and in-medium phase shift data.  If such
data is not available, an estimate can be obtained by using the corresponding
vacuum data, although as is shown in figure \ref {Deltafig} the difference 
could be significant.

In figure \ref{Deltafig} we display the resulting gap as a function of Fermi
momentum $p_F$, obtained by inserting $^1S_0$ phase shifts obtained
from the Nijmegen partial wave analysis of NN scattering data \cite{Nij}.
The result of (\ref{gapdelta}), using the bare neutron mass and zero-density
(non-medium) phase shifts, 
is given by the solid curve. Inserting the effective mass and
scattering amplitudes from \cite{Wam} yields the short-dashed curve.
The long-dashed and dot-dashed curves give the results of \cite{Wam} 
(lower curve) and \cite{Khodel} (upper curve).

\epsfysize=6 cm

It appears that the result (\ref{gapdelta}) 
is accurate as well as simple. Our results agree reasonably
well with 
state of the art calculations using more traditional methods. Our
leading result (ignoring medium corrections) 
is intermediate between calculations with ``induced potential''
effects \cite{Wam,Clark,Schulze}, which have a maximum gap size of
order 1 MeV, and those using a bare potential \cite{Schulze,barep,Khodel},
which produce a maximum gap of approximately 3 MeV. When medium effects
are included the gap is decreased significantly.


\bigskip
\noindent 
The authors would like to thank  \O. Elgar\o y  and C. Pethick
for useful correspondence concerning the low density limit.
Nick Evans, Stephane Keller and Myck Schwetz are acknowledged for
useful comments or discussions.
This work was supported in part under DOE contracts DE-AC02-ERU3075
and DE-FG06-85ER40224.


\end{document}